\documentclass[prd,aps,floats,showpacs,twocolumn,twoside,preprintnumbers,
superscriptaddress,floatfix]{revtex4}
\usepackage{bm}
\usepackage[T1]{fontenc}
\usepackage[latin1]{inputenc}
\usepackage{amsmath}
\usepackage{amssymb}
\usepackage{mathrsfs}
\usepackage{epsfig}

\newcommand{\bea}{\begin{eqnarray}}
\newcommand{\eea}{\end{eqnarray}}
\newcommand{\be}{\begin{equation}}
\newcommand{\ee}{\end{equation}}

\newcommand{\nnc}{\nonumber  \\
\nonumber  \\}
\newcommand{\nn}{\nonumber}

\newcommand{\bwt}{\begin{widetext}}
\newcommand{\ewt}{\end{widetext}}
		
\begin{document}

\title{Distribution of Matter from Singularity with Spherical Symmetry Using Fick Diffusion}

\vskip 1cm
\author{ Ahmet Mecit \"Ozta\c{s}}
\email{oztas@hacettepe.edu.tr}
\affiliation{Department of Physics, Hacettepe University,
TR-06800 Ankara, Turkey}
\author{Michael L. Smith}
\email{mlsmith10@cox.net}
\affiliation{Anabolic Laboratories, Inc. Tempe, AZ 85281, USA}

\begin{abstract}A model is presented allowing calculation of energy and matter distribution in the Universe after expansion from singularity without introduction of expansion energy. Beginning with Fick's law of diffusion, we solve the Bessel function for spherical systems during expansion, presuming isotropic matter distribution in the Hubble flow. This function can be substituted with the Associated Legendre differential equation and when solved we discover useful parameters for 
those of energy and matter densities, diffusion and temperature. Though we can follow the decline of matter density over spacetime, we cannot suggest a precise value at singularity. This model may be useful though, as a starting point for modeling star formation rates, SFR, galaxy formation horizon lengths and regions devoid of matter where tracing the matter density decline over time is important.
\end{abstract}
\pacs{04.20.Dw, 04.20.Ex, 95.30.Sf, 98.80.Jk,98.80.Bp}
\maketitle

\section{INTRODUCTION}
A prediction of the General Relativity \cite {Einstein:1952}, is the presence of singularity in our past. It seems all energy, matter and spacetime had once been confined within an incredibly small point which is now a mathematical and physical problem. While high energy experiments are helping unravel many details of particle formation soon after singularity \cite {Schwarz:2003}, cosmology might propose estimates for initial energy densities, 
particle and energy diffusion, temperatures, the nature of immature spacetime, \cite{Guth:1981}, \cite{Linde:1982}, \cite{La:1989} Universe age, initial galaxy size and star formaton rates (SFR). 
\vskip 0.2cm
A difficult problem of cosmology is the description of energy and particle distribution in spacetime after release of the Universe from singularity. Evidence from the cosmic microwave background (CMB), observed from a time about 380,000 years after singularity, suggests that energy distribution was extremely homogeneous \cite{Sievers:2003} and recent estimates of the smoothness of this radiation present a Gaussian distribution of very small fluctuations \cite{Chen:2005}. We also know that soon afterwards matter recombined into atomic H and He, quickly followed by the birth of dense stars, star clusters, galaxies and even black holes \cite{Dietrich:2001}. This early Universe, dense with matter, is thought to have been a violent era when huge stars devoured enormous amounts of matter ending "life" with incredible supernova (SN) explosions, littering the surrounding spacetime with many of the heavier elements we presently enjoy \cite{Kitayama:2005}\cite{Wise:2005}. This quick appearance of star-rich galaxies is puzzling and leads to suggestions that galaxy formation began before recombination \cite{Gibson:2000}.
\vskip 0.2cm
Much effort is being made in understanding the nature of galaxy and SFR by modeling matter distribution after recombination \cite{Springel:2005}. These semi-emperical models are quite detailed and are usually based upon cold dark matter (CDM) and dark energy considerations ($\wedge$\/CDM). These models indicate a maximum SFR between 11 and 13 Gyr ago, very soon after recombination \cite{Hernquist:2003}. SFR models are first order dependent upon matter density decline and gas temperatures, along with several truely empirical parameters. Unfortunately, SFR also seems dependent upon the nature of gathered data; dust contributes mightily to noise in the optical spectrum which complicates checking the models against relible data. Comparisons with radiotelescopic data have been made and the expected redshift dependence of SFR is observed but, significant observational technique dependence of data has also been reported \cite{Georgakakis:2006}.  Enourmous dependence is also placed upon cold dark matter (CDM) which is generally accepted but experimentally unconfirmed; perhaps existing as plentiful amounts of really exotic particles. CDM is thought to contribute much more to galaxtic gravity than shining matter, star remains, dwarfs and planets \cite{White:1993}. Dark energy may also play a role but is mathematically inconsistent and has changed from a repulsive to attractive force looking back to about z$\approx $\/5 \cite{Oztas:2006} an important epoch for SFR.
\vskip 0.2cm
We have generalized the first moments of the Universe using some well-known laws to describe matter and energy distributions back to near the beginning of time. We have used the general diffusion equation of Fick, assuming a single point was divisible into a nearly infinite numbers of small and equally proportioned components, to solve early spacetime expansion. The isotropic dispersion of matter allows assumpton of homogenous and isotropic, non-relativistic diffusion of identical and nearly innumerable components and we do not distinquish between ordinary matter and CDM. We have solved this specific situation as radial spacetime and by using the Galilean metric for matter and coupled radiation, have been able to predict regions of high matter density and those devoid of matter. This model is useful for predictions of matter density decline and for solutions of galaxy formation, very old SN rates and SFR. 
\vskip 0.2cm
\section{THEORY}
\subsection{Diffusion Model of the Universe}
The general diffusion equation of Fick can be applied to the Universe as matter diffusion 
from Singularity to present
\bea
\frac{\partial \rho}{\partial t}=\kappa\nabla^2\rho  +\mathbf{v}.\nabla  \rho
\label{difeq}
\eea
where $\rho$ is matter density and $\kappa$ is a diffusion constant and $\mathbf{v}$ is the expansion rate.

This equation is well known in fluid mechanics as the advection-diffusion equations for diffusion processes. We apply the Galilean transformation to particle dynamics in the Universe, after the cessation of inflation; we suggest particle flights are best modeled as non-relativistic with radiant energy strongly coupled to matter both diffusing well below light speed. We will apply a Galilean transformation to simplify the non-relativistic expansion by rewriting diffusion in a moving coordinate frame. Our Galilean frame is a radial expansion with a radial velocity not necessarily identical to the Hubble flow
\be
\mathbf{r}=\mathbf{{r^\prime}}-\mathbf{v} t  \quad\textrm{and}\quad
t'=t \nnc
\ee
and with the derivatives$^{[1]}$\footnotetext[1]{$\frac{\partial}{\partial t}=\frac{\partial t'}{\partial t}\frac{\partial}{\partial t'}+\frac{\partial\mathbf{r}'}{\partial t}.\nabla_{\mathbf{{r^\prime}}} \\
\hspace*{.35cm}=\frac{\partial}{\partial t'}+\mathbf{v}.\nabla_{\mathbf{{r^\prime}}}$}
\be
\displaystyle \nabla_{\mathbf{{r^\prime}}}= \nabla_{\mathbf{r}} \displaystyle{\frac{\partial}{\partial t}=\frac{\partial }{\partial t'}+\mathbf{v}.\nabla_{\mathbf{{r^\prime}}}}
\label{Galtr}
\ee
Placing these into Eq. (\ref{difeq}) we get the density dependence upon the new time and radial distance scales
\bea
\frac{\partial \rho}{\partial t'}+\mathbf{v}.\nabla_{\mathbf{{r^\prime}}} \rho&=&\kappa\nabla_{\mathbf{{r^\prime}}}^2\rho  +\mathbf{v}.\nabla_{\mathbf{{r^\prime}}}  \rho \nnc
\frac{\partial \rho}{\partial t'}&=&\kappa\nabla_{\mathbf{{r^\prime}}}^2\rho .
\label{difeq1}
\eea
For the diffusion of matter and spacetime from a spherical point into expanding spacetime let us suggest
that $\rho$ be a function of 3-dimensions and the variable temperature, which is itself a function of time, $\rho=u({r^\prime},{\theta^\prime},{\phi^\prime})T(t')$. This model for matter diffusion is the partial differential equation
\bea
u\frac{\partial T}{\partial t'}=T\kappa\nabla_{\mathbf{{r^\prime}}}^2u.\nnc
\eea
We now rearrange to separate the variables
\bea
\frac{1}{T}\frac{dT}{d t'}=\frac{1}{u}\kappa\nabla_{\mathbf{{r^\prime}}}^2u . \nn
\eea
Because the left and right hand sides are equal, both sides must also be equal to 
a constant 
\bea
\frac{1}{T}\frac{d T}{d t'}=\frac{1}{u}\kappa\nabla_{\mathbf{{r^\prime}}}^2u = -\lambda^2 
\label{sepvareq1}
\eea
which allows us the freedom of two independent equations
\bea
\frac{d T}{d t'}=-\lambda^2 T,\quad \kappa\nabla_{\mathbf{{r^\prime}}}^2u =-\lambda^2 u. 
\label{matterHubble}
\eea
For subsequent presentation we will designate these equations as "thermal relaxation" and "matter diffusion", respectively. 
We will now let $t$ represent some function of time, so the solution of the "thermal relaxation" model is 
$$T(t')=T_0 e^{-\lambda^2 t'}.$$
This model seems straight-forward as a grossly general description for times after cessation of inflation, but must be modified for the special conditions during that epoch.

To begin the solution of the "matter diffusion" portion we will conform to the usual conventions where the $\nabla^2$ operator is the Laplacian and the $\nabla$, del operator, is the total partial operator. Using these terms the expansion has the following forms in spherical coordinates
\bea
{\nabla_{\mathbf{{r^\prime}}}}^2&=&\frac{1}{{r^\prime}^2}\frac{\partial}{\partial {r^\prime}}
\left({{r^\prime}}^2\frac{\partial }{\partial {r^\prime}}\right)-\frac{L^2}{{{r^\prime}}^2}  \\ 
L^2&=& -\left(\frac{1}{\sin{\theta^\prime}}\frac{\partial}{\partial {\theta^\prime}}\left(\sin{\theta^\prime}\frac{\partial}{\partial {\theta^\prime}}\right)+\frac{1}{\sin^2{\theta^\prime}}\frac{\partial^2}{\partial {\phi^\prime}^2}\right).
\eea

We will now allow $\dot{r}=v_{r}, $ while assuming that $\dot{\theta}=0$ and $\dot{\phi}=0$ in Eq. (\ref{Galtr}). That is, the averaged polar and azimuthal velocities of matter are zero when traveling in the new reference frame, consistent with the presumption of a homogeneous dispersion of particles beginning with $t$ = 0.  Such is also consistent with the very homogeneous nature of the CMB radiation. Since this operator involves angular and radial variables we shall separate each variable for a useful solution. The solutions of the angular portions of the operators are the angular functions $Y_{l,m}({\theta^\prime},{\phi^\prime})$ presented in Appendix A and are well-known in quantum mechanics. To use the step function

\be L^2Y_{l,m}(\theta,\phi)=l(l+1)Y_{l,m}(\theta,\phi)\label{am}\ee

\noindent we suggest $u({r^\prime},{\theta^\prime},{\phi^\prime})=R({r^\prime})Y_{l,m}({\theta^\prime},{\phi^\prime})$ for the "matter diffusion" portion of equation \ref{matterHubble}.
\bea 
\kappa\nabla_{{r^\prime}}^2u =-\lambda^2 u.  \nn
\eea
We will separate $u$ into radial and angular dependencies, which in a compact for is

\bea
\frac{1}{{r^\prime}^2}\frac{\partial}{\partial {r^\prime}}\left({r^\prime}^2\frac{\partial }{\partial {r^\prime}}
\right)u+\frac{L^2}{{r^\prime}^2}u=-\frac{\lambda^2}{\kappa} u.  \nn
\eea

We can now rewrite this equation after separation of variables and substituting with Eq. (\ref{am}) as

\be
{r^\prime}^2\frac{d^2R({r^\prime}) }{d {r^\prime}^2}+2{r^\prime}\frac{dR({r^\prime}) }{d {r^\prime}}+\left(k^2{r^\prime}^2
-l(l+1)\right)R(r^\prime)=0
\label{parteq1}
\ee
where $k^2=\frac{\lambda^2}{\kappa}$, remembering that ${\kappa}$ is the diffusion constant.

Let us now substitute for $y=k{r^\prime}$ 
and by placing $y$ into Eq.(\ref{parteq1}) we have the following equation
\bea
y^2\frac{d^2R(y)}{d y^2}+2y\frac{dR(y)}{d y}+\left(y^2-l(l+1)\right)R(y)&=&0.
\label{parteq2}
\eea
We will now use a second substitution of $ \displaystyle R(y)=Z(y)/\sqrt{y} $. 
When we use this for $R(y)$ in Eq. (\ref{parteq2}) we are left with a
solution which is in the form of the Bessel differential equation of the type
\bea
y^2\frac{d^2Z(y)}{d y^2}+y\frac{dZ(y)}{d y}+(y^2-(l+\frac{1}{2})^2)Z(y)=0.
\label{rdep1}
\eea

Two solutions of which are essentially spherical Bessel functions(some primitive solutions are presented in Appendix B) which we now define as
$$ j_l(k{r^\prime})=\sqrt{\frac{\pi}{2}} \frac{J_{l+1/2}(kr^\prime)}{\sqrt{k{r^\prime}}}, \qquad n_l(k{r^\prime})=\sqrt
{\frac{\pi}{2}} \frac{Y_{l+1/2}(kr^\prime)}{\sqrt{k{r^\prime}}}.$$
The physical solution for both of these is well-behaved; $ j_l(k{r^\prime})$, so the solution of radial portion is the proportionality
\be
R_l({r^\prime})\propto j_l(k{r^\prime})
\label{solr}
\ee
and the general solution of the Eq.(\ref{difeq1}) is now
\bea
\rho({r^\prime},\theta,\phi;t) 
&=&\sum_{l=0}^{\infty}\sum_{m=-l}^lA_{l,m}e^{-\lambda_{l,m}^2 t}j_l(k_{lm}{r^\prime})
Y_l^m(\theta,\phi).~~~~~~
\label{final}
\eea

Where we allow $\phi^\prime=\phi$, and $\theta^\prime=\theta$, because assume the expansion from singularity behaves isotropically and has only the radial component as non-isotropic. The path to the final result below is presented in Appendix C, where the orthogonal solutions to the Bessel functions are presented
\bea
A_{l,m}=\frac{2 k_{lm}^2}{\pi}\int_Vj_l({k_{lm}{r^\prime})}Y_{l}^{m}(\theta,\phi)\rho({r^\prime},
\theta,\phi;0) d^3r^\prime
\label{result}
\eea

\subsection{Substitution of the Hubble constant for $\lambda$}

\noindent If one recalls $r=r_0 a(t) $ where $r_0$ is the present Universe radius, and $a(t)$ is the expansion parameter (Hubble's law) is
\be v=\frac{d r}{dt}=r_0\frac{d a(t)}{dt}=Hr. \ee

\noindent Dividing each side r, inserting the matter density $\rho=M/(4\pi r^3/3)$ then taking the derivative we have
\bea\frac{d \rho}{dt}&=& -\frac{3 M} {(4 \pi r^4/3)} \frac{d r}{ d t} \nnc
\frac{ d \rho} {dt}&=&-3 \rho \frac{1}{r} \frac{d r}{ d t}. \nn\eea

\noindent Then dividing by each side with $\rho$ and inserting the Hubble Constant

$$\frac{1}{\rho }\frac{d \rho}{ dt} =-3 H. $$
\noindent One can derive the left hand side from our Eqs. \ref{Ex2} , \ref{Ex3b}  or \ref{exampdirac}

$$\frac{1}{\rho}\frac{ d \rho}{dt} =-\lambda^2$$
\noindent 
so that we arrive at a compact form for our constant connecting temperature decline with matter diffusion as 
\be \lambda^2=3 H.
\label{lamhub}
\ee
At some time in the past, the Galilean frame becomes unnecessary as small perturbations merge togther and the Universe is better described by the Hubble expansion. 
\subsection{Path of light in the early Einstein Universe}
\par\noindent
We suggest much of the energy of diffusion in the expanding Universe just after recombination can be considered uncoupled photons.  These "excess" photons were subject to the conditions of the Einstein Universe and can be treated with the usual metric
\[ds^{2} = -(1-\frac{r^{2}}{\mathbb{R}^{2}})^{-1}dr^{2} - r^{2}d\theta ^{2} - r^{2} \sin ^{2}\theta d\phi ^{2} + c^{2}dt^{2}.\]
\par\noindent
The path of a photon, if collision free during the initial moments of recombination, will be subject to spacetime curvature in the usual manner with $\mathbb{R}$\/ a constant of spacetime curvature.  In a Universe with truly homogeneously and isotropically dispersed matter and energy we should find $\mathbb{R}$\/ to be quite large and the light path to be nearly, if not completely flat, even in the primitive Universe.  We expect inhomogeneties in the polar and azimuthal dimensions were extremely small, contributing only very slightly to the curvature of photon flight. Light in a primitive Universe without truly isotropically dispersed matter and energy may suffer a flight through curved spacetime and a slightly smaller value for $\mathbb{R}$\/. Such radiation might be observed at later times as our CMB \cite{Stellmach:1999} with much of the tiny temperature ripples due to more recent interactions with matter and supernova additions \cite{Kitayama:2005}, but still containing information before recombination when more recent events are removed \cite{Chen:2005}.

\section{Examples of Conditional Solutions}
The following examples might be considered models for our Universe approaching $t=0$. These are based on a function describing a declining proportionality between matter density in spacetime and the time/radial distance from the origin. The reader should remember that $\lambda ^{2} = 3H$ in the following examples which are based upon the proportionality 
$$\rho(r,\theta,\phi;t) \varpropto \frac{\sin(k(r+vt))}{r+vt}e^{-3 H t}$$
The parameters $H$ and $k$ may be chosen from observations and we should also remember that $k^{2}\varpropto\frac{1}{\kappa}$; the square of the matter density is inversely proportional to a diffusion constant.
\label{Examples}
\subsection{Spherical Symmetric Initial Matter Density}

With commencement of the Universe, matter distribution was approximately isotropic and can be represented as $\rho(r^\prime,\theta,\phi;0)=\rho(r^\prime)$. In this case Eq. (\ref{result}) can be reduced to only the radial integral with the help of relationships in Appendix A, Eq. (\ref{ylm0})
\bea
A_{l,m}\!&=&\!\frac{4 k_{lm}^2}{\sqrt{\pi}}\delta_{l,0}\delta_{m,0}\int_0^\infty {r^\prime}^2\rho({r^\prime}) j_l(k_
{lm}{r^\prime}) d{r^\prime} 
\label{sphalm}
\eea

In this example we have choosen a total mass of incredible density, $\rho(r^\prime,
\theta,\phi;0)\sim 1/({r^\prime}^n)$. By substituting into Eq. (\ref{sphalm}) and attempting to calculate the initial matter density we find the result is highly dependent both on the radial distance and time approaching the Universe origin.  The total for matter and energy is represented by C in the relationship below
\bea
A_{l,m}\!&=&C\frac{4 k_{lm}^2}{\sqrt{\pi}}\delta_{l,0}\delta_{m,0}\int_0^\infty {r^\prime}^2\frac{1}{{r^\prime}^n} j_l(k_
{lm}{r^\prime}) d{r^\prime} \nn
\eea

where $y=k_{lm}{r^\prime}$. This function may be solved using a suggestion from \cite{Abramowitz:a}, by allowing $\mu=3/2-n$ and $\nu=l+1/2$ for a solution of the integral for the radial distribution in Eq. (\ref{result}) as
\be 
\int_0^\infty t^\mu J_\nu(t)dt=\frac{2^\mu\Gamma(\frac{\mu+\nu+1}{2})}{\Gamma
(\frac{\nu-\mu+1}{2})}
\label{xpj}
\ee
and with substiution for the integral becomes
\be
A_{l,m}=4C\left(\frac{k_{lm}}{2}\right)^{n-1}\delta_{l,0}\delta_{m,0}\frac{\Gamma(\frac{l-n+3}
{2})}{\Gamma(\frac{l+n}{2})}.
\ee
By substituting this into Eq. (\ref{final}) and using $\delta$ functions we may generate the general case for the initial asymptotic appearance of energy and matter density in our Universe

\bea
\rho(r^\prime,\theta,\phi;t)&=& 
4C\left(\frac{k}{2}\right)^{n-1}\frac{\Gamma(\frac{3-n}
{2})}{\Gamma(\frac{n}{2})}
\frac{\sin(kr)}{kr}e^{-3 H t} Y_{0}^0(\theta,\phi)\nnc
&=&2C{\frac {\sin \left(\frac{n\pi}{2} \right) \Gamma  \left( 2
-n \right) }{ \pi k^{1-n} }}\frac{\sin(kr^\prime)}{kr^\prime}e^{-3 H t}.
\label{SSrmn}
\eea
It seems that Eq. (\ref{SSrmn}) will be difficult to evaluate since the value of the pre-exponential portion is highly dependent upon the relative value of the radial distance as our lookback approaches singularity, while the exponential portion, dependent upon Hubble expansion, is time dependent. One thing to note from the pre-exponential factor is that energy/matter density depends on the diffusion of both through spacetime, as reflected by $k$.  This is also consistent with the properties of a gravitational field, for particle distribution in any gravitational field is dependent upon radial distances between objects.

\subsubsection{Homogeneous and Constant Matter Density at $t=0$ and\\ $n=0$}
Let us presume that the simplest case for equation (\ref{SSrmn}), $n=0$, corresponds to a constant initial energy/matter density $\rho(r^\prime,\theta,\phi;0) = C/{r^\prime}^0=C$, and that energy/matter was homogeneously distributed at singularity, $t=0$. We can now use Eq. (\ref{SSrmn}) by allowing $\rho(r^\prime,\theta,\phi;0)=C=$constant. 
\be
\rho(r^\prime,\theta,\phi;t) 
=2C{\frac {\sin \left(0 \right) \Gamma  \left( 2
\right) }{ \pi k }}\frac{\sin(k{r^\prime})}{k{r^\prime}}e^{-3 H t}. \nn
\ee
Due to $\sin(0)=0$ we are left with
\bea
\rho(r^\prime,\theta,\phi;t)&=&0.\label{Ex1}
\eea
This result seems both puzzling and logical; suggesting that matter density was not distributed at all outside singularity. This also suggests that homogeneously distributed matter at singularity was impossible, at least with this model. Perhaps at the commencing of the Universe, $t=0$, constant matter density everywhere within Universe cannot be well described, except to suggest that matter density was 0 everywhere else but at singularity, whatever that "else" was.  Another interpretation, though contrary to common thought, might be that at the instant of singularity there was intially no energy or matter density, at all. This suggestion, though strange, may also be a valid estimate of singularity at origin.
\newpage
\subsubsection{The Matter Density Asymptote at $t=0$  and\\$0<n\leq2$}
We again borrow from \cite{Abramowitz:b}, $\Gamma(\frac{1}{2})=\sqrt{\pi}$ and $k_{0,0}=k, \lambda_{0,0}=\sqrt{3 H}$,
and  we shall evaluate our model at $n=1$ and $n=2$ but not $n=0$ in Equation (\ref{SSrmn}) 
\be
\rho(r^\prime,\theta,\phi;t) 
=
\left\{\begin{array}{ll}
\displaystyle C\frac{2}{\pi}\frac{\sin(k{r^\prime})}{k {r^\prime}}e^{-3 H t},&n=1 \\
\\
\displaystyle C\frac{\sin(k{r^\prime})}{{r^\prime}}e^{-3 H t},&n=2
\label{Ex2}
\end{array}
\right.
\ee
\label{n=1,2}
Here we have arrived at an estimate of the initial matter density of the Universe approaching singularity with some real values for the pre-exponential portion of our model. As expected from logical considerations and the arguments of \cite{Hawking:1973} the total energy/matter is still incalculable, except to suggest it approaches infinite.
\subsubsection{An example of the above with the boundary condition:  $n >2, n=$even integers}

Let us now presume for $n>2=2m$, $m=2,3,4,...$, by again substituting real energy/matter density into Eq. (\ref{SSrmn}) $\rho({r^\prime},\theta,\phi;0) = C/{r^\prime}^{2m}$ and 
\bea
\rho(r^\prime,\theta,\phi;t) &=&2C{\frac {\sin \left(m\pi \right) \Gamma  \left( 2
-2m \right) }{ \pi k^{1-2m} }}\frac{\sin(k{r^\prime})}{k{r^\prime}}e^{-3 H t}\nnc
&=&\frac{(-1)^{m-1}k^{2m-1}}{\Gamma(2m-1)}\frac{\sin(k{r^\prime})}{k{r^\prime}}e^{-3 H t}
\label{Ex3b}.
\eea

Here, the very left-hand portion of the relationship is again finite and the exponential a function of $t$ similar to the previous example of $n=1,2$. The complete pre-exponential cannot be evaluated at $r=0$ though. The above solution may be useful as a model, for at large times, radial distances and with a realistic diffusion constant the matter density can be calculated.  As time and radial distance approach the infinite, the energy/matter density declines to 0, one predicted fate of our Universe.
\subsubsection{An example of a singularity created with the boundary condition:  $n>2,n=$odd integer}

Let us presume for $n=2m+1$, $m=1,2...$, and with substitution of real matter density into Eq. (\ref{SSrmn}) that $\rho({r^\prime},\theta,\phi;0) = C/{r^\prime}^{2m+1}$. If we develop this Universe along the manner of previous examples, spherical symmetry but with spacetime in the forward looking direction, we can evaluate once again
\bea
\rho(r^\prime,\theta,\phi;t) &=&2C{\frac {\sin \left(m\pi+\frac{\pi}{2} \right) \Gamma  \left( 1-2m \right) }{ \pi k^{-2m} }}\frac{\sin(k{r^\prime})}{k{r^\prime}}e^{-3 H t} \nnc
&=&2C{\frac {(-1)^m \Gamma  \left(1 -2m \right) }{ \pi k^{-2m} }}\frac{\sin(k{r^\prime})}{k{r^\prime}}e^{-3 H t}\nnc
&=&\infty
\label{Ex3a}
\eea
This means the energy/matter density, everywhere and at all times, is always infinite. This might be interesting but not believable.

\subsection{Solutions of a Point-Like Universe at $t=0$}
\subsubsection{Squeezed in a little sphere of $r=a$}
We again examine the Universe at $t=0$ and allow the matter density 
to be a smooth function of the spherical spacetime density with
$\rho~\sim  \Theta(a-r) $ , where $ \Theta(a-r) $ is the Heaviside, 
step function. If $a-r>0$ it obtains the value of 1, otherwise it is  0, which 
truncates the $r$ integral at $r=a$.  So we will examine a sphere with radius of $a$, 
where the density, $C$, may be equal to $M/(4\pi a^3/3)$. Let us suggest, for this solution, 
that $\rho(r^\prime,\theta,\phi;0)=C\Theta(a-{r^\prime})$ and substitute this into spherical symmetric case Eq. (\ref{sphalm}) then evaluate over all space

\bea
A_{l,m}\!&=&\!\frac{4 k_{lm}^2}{\sqrt{\pi}}\delta_{l,0}\delta_{m,0}\int_0^\infty {r^\prime}^2\rho({r^\prime}) j_l(k_
{lm}{r^\prime}) d{r^\prime}  \nnc
&=&C\frac{4 k_{lm}^2}{\sqrt{\pi}}\delta_{l,0}\delta_{m,0}\int_0^\infty {r^\prime}^2 \Theta(a-{r^\prime})j_l(k_
{lm}{r^\prime}) d{r^\prime}  \nnc
&=&\frac{M}{(4\pi a^3/3)}\frac{4 k_{lm}^2}{\sqrt{\pi}}\delta_{l,0}\delta_{m,0}\int_0^a {r^\prime}^2 j_l(k_{lm}{r^\prime})
d{r^\prime}\nnc
&=&\frac{3 M k}{\pi^{3/2} a^3}\delta_{l,0}\delta_{m,0}\left(\frac{\sin(ka)}{k^2}-
\frac{a\cos(ka)}{k}\right). \nn
\eea
Further progess may be made by evaluating the series expansion for small values of $a$ which obviously demands large initial matter densities 
\be
A_{l,m}\approx\frac{3 M}{\pi^{3/2} a^3}\delta_{l,0}\delta_{m,0}\left(\frac{1}{3}k^{2}-\frac{1}{30}k^{4}{a}^{2}+O \left( {a}^{3} \right) \right).
\ee

After substitution into Eq. (\ref{final}) and evaluation of the $\delta$ functions the result, in terms of the Hubble constant, is dependent on the familiar parameters of time, radial distance, matter and diffusion
\bea
\rho(r^\prime,\theta,\phi;t) 
&=&\frac{ M k^2}{2 \pi^2 }\frac{\sin(k{r^\prime})}{k{r^\prime}}e^{-3 H t}
\label{exampspher}
\eea
For the solution leading to Eq. (\ref{exampspher}) we used the first term of the expansion when we took the limit of $A_{l,m}$ at $a=0$ for the shrinking sphere. The solution of Eq. (\ref{exampspher}) is infinite initial matter density at singularity, in as compact and precise form as can be presently suggested. This solution is similar to our Eq. (\ref{Ex3b}) where the matter density approaches infinite as the diameter of the Universe approaches 0.  So we have obtained similar results from two differing initial conditions.
\label{littlesphere}
\subsubsection{Point-like in Radial Direction Using the Dirac Function}
Imagine the Universe at singularity being compressed to a point with a 
volume of zero. It is possible to represent this mathematically by a 
special function; the Dirac delta function and we shall use this to represent 
a tiny particle by a single point. Let us assume at $t=0$ that $\displaystyle \rho(r^\prime,\theta,\phi;0)=C 
\frac{1}{{r^\prime}^2\sin\theta}\delta(r^\prime)$, where we do indeed have real energy/matter density at singularity. 
We have substituted this condition into Eq. (\ref{result}) and have evaluated as follows
\bea
A_{l,m}&=&\frac{2 k_{lm}^2}{\pi}\int_Vj_l({k_{lm}{r^\prime})}Y_{l}^{m}(\theta,\phi)\rho
(r^\prime,\theta,\phi;0) d^3{r^\prime} \nnc
&=&C\frac{2 k_{lm}^2}{\pi}\int_0^\infty j_l(k_{lm}{r^\prime})\delta({r^\prime})d{r^\prime} \nnc
&&\times\int_0^{2\pi}\int_0^\pi Y_{l}^{m}(\theta,\phi)  d\theta d \phi \nnc
&=&C\frac{2 k_{lm}^2}{\pi}\sqrt{\pi} j_l(0)\sqrt{{2l+1}}
\delta_{m,0}\left(\frac{\Gamma(l+1/2)}{l!}\right)^2\nnc
&=&C\frac{2 k_{00}^2}{\pi}\sqrt{\pi} \delta_{l,0}
\delta_{m,0}(\sqrt{\pi})^2 \nnc
&=&2C\sqrt{\pi}k^2\delta_{l,0}\delta_{m,0}.
\eea
The angular integral of the spherical harmonics can be evaluated as in Appendix A Eq. (\ref{intYlm}). Here, we may use real values for matter density, diffusion and the expansion rate of the Universe; $j_l(0)
=\delta_{l,0}$   \cite{Abramowitz:c} and substitute into Eq. (\ref{final}) and again using $\delta$ functions
\bea
\rho({r^\prime},\theta,\phi;t) 
=\rho_0 \frac{\sin(k{r^\prime})}{k {r^\prime}} e^{-3 H t}.
\label{exampdirac}
\eea
\label{Dirac}
We now have another result of interest, arrived by taking a separate path, for at most any radial distance from singularity, given the diffusion constant, time and expansion rate, we should be able to calculate the matter density of the Universe, in a seemingly straightforward manner, using Eqs. (\ref {exampspher}, \ref {exampdirac}).

\section{DISCUSSION}
We have presented a model for the distribution of matter in the Universe beginning at singularity with dependence on time, Hubble flow, and initial matter density. By choosing as our initial condition the coupling of energy/matter density with thermal relaxation and Galilean comoving coordinates, we have been able to solve spacetime expansion using Bessel's spherical functions. To do this we first presumed expansion in the radial direction away from singularity but as spacetime expansion progresses the radial dimension degenerates into the present situation of relative spacetime. By separating Eq.(\ref{sepvareq1}) into two equal portions we have also assumed a constant value for total energy and matter immediately after singularity. We have chosen to not introduce continuous synthesis of energy after singularity, even with a term similar to that first presented by Einstein \cite{Einstein:1952} and of current interest \cite{Reiss:2004}; this tends to cloud straight-forward evaluation because some solutions display regions of discontinuity \cite{Oztas:2006}. 

We have used our solutions to examine several models for the Universe close to singularity. Such tests are important because models of this epoch must be inferred from radiation emitted at much later times, which has been modified by intervening epochs \cite {Kitayama:2005}, and current high energy experiments. The examples from our model are consistent, to the first order, with current CMB data by being homogeneous and angular independent. 

In Sec. \ref{Examples} we attempted to calculate energy/matter density back to singularity, that is towards $t=0$. We presumed the initial matter density as spherically symmetric and used $r^{-n}$ type functions with asymptotes at $r=0$. This may be the simplest model one can use to describe singularity with respect to energy/matter density. The general solution of these type of functions, Eq. (\ref{Ex3b}) may seem straightforward, but insertion of specific values for $n$, for instance, selecting either even or odd values for $n>2$, leads to differing results depending on the pre-exponential factors, which in turn depend on the integer chosen for the solution. An example solution taking this approach is (\ref{Ex1}), which is the special case of $n=0$, appearing not radially homogeneous, a condition we think is required by current CMB data, but we suspect this means everything but singularity exhibited no energy/matter density - at least to an "outside" observer. Another example (\ref{Ex3a}) describes matter density as infinite everywhere at real values for the radius of the Universe, so this model is not useful.

We have found, through example \ref{n=1,2}, the $n=1,2$ case, and Eqs. (\ref{Ex2}) that we may only estimate the matter density at singularity as incredibly large as the Universe dimensions approach 0. We have found that matter density close to singularity is also very dependent on matter diffusion, as expressed with the parameter, $\kappa$ again in Eqs. (\ref{Ex2}). Examining these equations at the extreme situation of no matter, our model is consistent with that of DeSitter, where spacetime is absolutely flat and infinite, for when $\kappa = \infty $ all matter must disappear.  So this model, like the DeSitter Universe, falls short of observations. When another solution, Eq. (\ref{Ex3a}), is evaluated, we find the Universe consisted of infinite matter density while displaying a non-zero radius. This seems impossible to us because there seems no reason for the eventual breaking of this "zone of infinite matter density" and allowing, spacetime expansion, thermal decline and matter diffusion. So this model is interesting but not useful.

Testing the examples in subsection \ref{littlesphere} for a Universe with diameter $r > 0$ we found that as the Universe is compressed towards singularity we obtain a solution, Eq. (\ref{exampspher}), with some similarity to that derived previously, Eq. (\ref{Ex3b}), but with a pre-exponential term we may possibly trace as close to singularity as we wish. For derivation (\ref{exampspher}), energy and matter densities are highly dependent on the radius of the Universe, the initial matter density, spacetime expansion and diffusion but will be incredibly large and nearly impossible to calculate accurately at singularity. When we use the Dirac $\delta$ function, in section \ref{Dirac}, we also derive a matter density, Eq. (\ref{exampdirac}), both similar to Eq. (\ref{exampspher}) and highly dependent on the parameters mentioned for that equation. These two solutions may allow us another avenue to estimate the initial energy/matter density of our Universe if we can estimate values for
the Hubble constant, matter diffusion and Universe age with accuracy near singularity. 
\vskip 0.2cm
We can present an interesting example of introducing a perturbation, which we shall denote as $\chi $\/, into Eq.(\ref{exampdirac}). We shall substitute with $\chi $\/ by using the proportionality introduced in Section III, as the Galilean metric.  If we let $\chi=kv/H $\/ we can recast equation (\ref{exampdirac}) in terms of the Hubble constant, time, distance and inversely proportional to a general matter difussion constant, compactly written as

\be \frac{\rho}{\rho_0}=\frac{\sin(kr+\chi Ht)}{(kr+\chi Ht)}e^{-3Ht} 
\label{comsol}
\ee
This equation allows us the flexibility that some fraction of matter diffuses with a slightly differing velocity than the Hubble flow. When we force $\chi $\/ to be zero, for the unperturbed case, we show in Figure 1 that we can follow the increase of the matter density with lookback time as a function of the $Ht$ while selecting for convenient values for the mean diffusion contant.  We cannot find evidence, from this plot, for an epoch of discontinuity or "jerk" as proposed by Reiss and coworkers \cite{Reiss:2004}; such must come from models using an explicit cosmological constant \cite{Oztas:2006}.  This situation of Fig. 1, also predicts epochs of negative matter density but only for special situations returning towards singularity. So this model appears much too general to indicate zones of lacking matter or rich in matter density.
\vskip 0.5cm
\begin{figure}[!ht]
\includegraphics[width=8.5 cm]{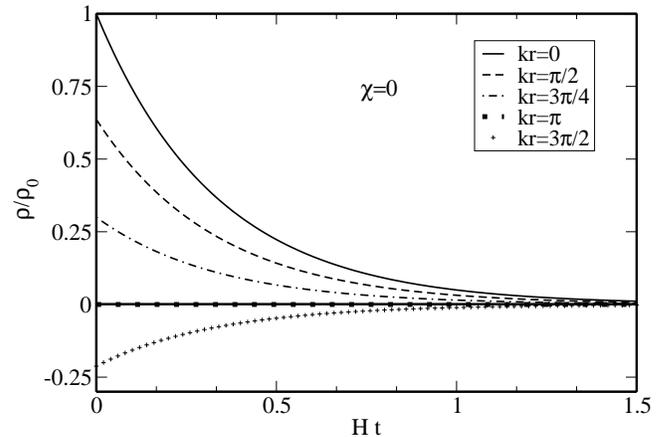}
\begin{center}
\caption{The behavior of the common solution from present matter density traced back towards early matter density in Eq. (\ref{comsol}) for selected values of $kr$ at $t=0$}\end{center}
\end{figure}
\vskip 0.5cm
We next display another case of this relationship as a surface in Figure 2, beginning at singularity with $\chi=5$\/ rather than 0.  Here we see that matter density decays very rapidly with increasing $Ht$ and that at significant times beyond singularity we see, by tracing regions close to the $kr$\/ axis, regions devoid of matter are predicted. Likewise, other regions become enriched in matter with increasing $Ht$ at the expense of neighboring spacetime.
\vskip 0.5cm
\begin{figure}[!ht]
\includegraphics[width=8.5 cm,height=7.5cm]{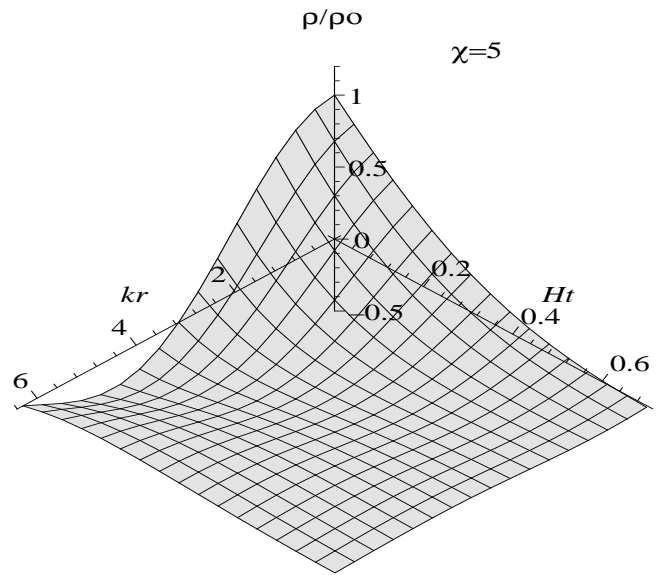}
\begin{center}
\caption{The behaviour of the common solution from present matter density traced back to the early matter density ($\rho/\rho_0$) versus $kr$ and $Ht$ for $\chi=5$}\end{center}
\end{figure}
The model displayed in Fig. 2 may be useful for estimates of the effects of perturbations within the primitive Universe, at non-relativitic matter velocities, effecting the horizon lengths which determine early galactic group volumes. Such perturbations, in more detailed forms, may be introduced into Eqs. (\ref{exampspher}) or (\ref{exampdirac}), for instance in the manner of Eq.(\ref{comsol}). Models including deviations from a Universe of ideal symmetry and including perhaps several Galilean metrics, may be found useful to reflect the small inhomogeneities, observed today, across the cosmic microwave background. Such perturbations, which may have been extremely tiny just after singularity, may well have become the nuclei for gas clouds, the first stars, clusters and galaxies.

\appendix

\section{Examples of Solutions to Spherical Harmonics}

The mixture of Associated Legendre and spherical functions can be presented in 
spherical harmonics as $$Y_l^m(\theta,\phi)=\displaystyle \sqrt{\frac{2l+1}
{4\pi}\frac{(l-m)!}{(l+m)!}}P_l^{m}(\cos\theta)e^{im\phi}.$$
Spherical harmonics condenses all of the terms, through 
the relation between Legendre functions and the range of $m$, $m=-l...l$

A few examples of the simplest solutions to spherical harmonics for $Y_l^m(\theta,\phi)$:

\bea
Y_0^0(\theta,\phi)&=&\frac{1}{2}\frac{1}{\sqrt{\pi}} \nnc
Y_1^0(\theta,\phi)&=&\frac{1}{2}\frac{3}{\sqrt{2\pi}}\cos \theta \nnc
Y_1^{\pm 1}(\theta,\phi)&=&\mp\frac{1}{2}\sqrt{\frac{3}{2\pi}}\sin \theta e^{\pm i\phi} \nnc
Y_2^0(\theta,\phi)&=&\frac{1}{4}\sqrt{\frac{5}{\pi}}(3\cos^2\theta-1)\nnc
Y_2^{\pm1}(\theta,\phi)&=&\mp\frac{1}{2}\sqrt{\frac{15}{2\pi}}\sin\theta\cos\theta e^{\pm i\phi} \nnc
&&\textrm{\vdots}
\eea

All of these solutions are related through conforming with the orthogonality relationship
\bea
\int_0^{2\pi}\int_0^\pi Y_l^m(\theta,\phi)Y_{l'}^{m'}(\theta,\phi)\sin\theta 
d\theta d\phi&=&\delta_{ll'}\delta_{mm'}~~~~~
\label{ortsphar}
\eea
\vskip 1cm
\bea
\int_0^{2\pi}\int_0^\pi Y_{l}^{m}(\theta,\phi) \sin\theta d\theta d \phi&=&2\sqrt{\pi}\int_0^{2\pi}\int_0^\pi\nnc &&\times Y_{l}^{m}(\theta,\phi)Y_{0}^{0}(\theta,\phi) \sin\theta d\theta d \phi\nnc
&=&2\sqrt{\pi}\delta_
{l,0}\delta_{m,0}
\label{ylm0}
\eea
\vskip 1cm
\bea
\int_0^{2\pi}\int_0^\pi Y_l^m(\theta,\phi) 
d\theta d\phi&=& \sqrt{\frac{2l+1}
{4\pi}\frac{(l-m)!}{(l+m)!}}\int_0^{2\pi}e^{im\phi}d\phi\nnc &&\times \int_0^\pi P_l^{m}(\cos\theta)d\theta
\nnc
&=& \sqrt{\frac{2l+1}
{4\pi}}2\pi\delta_{m,0} \int_0^\pi P_l(\cos\theta)d\theta
\nnc
&=& \sqrt{\frac{2l+1}
{4\pi}}2\pi\delta_{m,0} \int_{-1}^1\frac{P_l(x)}{\sqrt{1-x^2}}dx
\nnc
&=&\sqrt{\pi} \sqrt{{2l+1}}
\delta_{m,0}\left(\frac{\Gamma(l+1/2)}{l!}\right)^2\nnc
\label{intYlm}
\eea
The result of the integral of the Legendre polynomial is consistent with the formula of \cite{Gradsteyn:1981}

\section{Some solutions to the Spherical Bessel Functions}

Some of the primitive solutions to the spherical Bessel functions:

\bea
j_0(x)&=&\frac{\sin x}{x} \nnc 
n_0(x)&=&-\frac{\cos x}{x} \nnc 
j_1(x)&=&\frac{\sin x}{x^2}-\frac{\cos x}{x} \nnc
n_1(x)&=&-\frac{\sin x}{x}-\frac{\cos x}{x^2}\nnc
j_2(x)&=&\big(\frac{3}{x^3}-\frac{1}{x}\big) \sin x-\frac{3}{x^2}\cos x \nnc  
n_2(x)&=&-\frac{3}{x^2}\sin x-\big(\frac{3}{x^3}-\frac{1}{x}\big) \cos x\nnc
&&\textrm{\vdots}
\eea

\be 
\int_0^\infty r^2j_l(k_{lm}r)j_l(k_{lm}r)dr
=\frac{\pi}{2 k_{lm}^2}
\label{ortspbes}
\ee
This is based on a more general solution by Joachain \cite{Joachain}

\section{Calculating the Expansion Coefficients}

The  problem is not quite solved for we must perform further evaluations to uncover the coefficients. 
This requires numerical solutions for each $A_{l,m}$ for combinations of $l,m$. To do this we must impose 
orthogonality conditions upon Spherical Bessel solutions and spherical harmonic solutions; more details of 
these solutions are presented in Appendixes A and B. The general way to perform this is to multiply these 
solutions by orthogonal polynomials containing different parameters and then integrate over space, using 
the volume integral \\ $$\displaystyle \int_V...d^3{r^\prime}=\int_0^\infty\int_0^{2\pi}\int_0^\pi...{r^\prime}^2\sin\theta d\theta d \phi d{r^\prime} $$ 

A sample is given below if we reconsider  Eq. (\ref{final})

\bea
\rho(r^\prime,\theta,\phi;t) 
&=&\sum_{l=0}^{\infty}\sum_{m=-l}^lA_{l,m}e^{-\lambda_{l,m}^2 t}j_l(k_{lm}{r^\prime})
Y_l^m(\theta,\phi) \nn
\eea

to use  orthogonality relations we multiply each side $j_{l'm'}({k_{l'm'}r)}Y_{l'}^{m'}(\theta,\phi)$ and integrate over space and  then  then  denote  left hand side with $I$ of the equality
\bea
I&=&\int_Vj_{l'm'}({k_{l'm'}{r^\prime})}Y_{l'}^{m'}(\theta,\phi)\rho({r^\prime},\theta,\phi;t) d^3{r^\prime}\nnc
&=&\sum_{l=0}^{\infty}\sum_{m=-l}^lA_{l,m}e^{-\lambda_{l,m}^2 t}\int_Vj_{l'm'}
({k_{l'm'}{r^\prime})}Y_{l'}^{m'}(\theta,\phi)\nnc &&\times j_l(k_{lm}{r^\prime})Y_l^m(\theta,
\phi)d^3r\nnc
&=&\sum_{l=0}^{\infty}\sum_{m=-l}^lA_{l,m}e^{-\lambda_{l,m}^2 t}\int_0^\infty 
{r^\prime}^2j_l(k_{lm}r)j_l(k_{l'm'}{r^\prime})d{r^\prime} \nnc
&&\times \int_0^\pi\int_0^{2\pi}Y_l^m(\theta,\phi)Y_{l'}^{m'}(\theta,\phi)
\sin\theta d\theta d \phi \nn.
\eea

Eq. (\ref{ortspbes}) for the spherical Bessel solution and for spherical harmonics Eq. 
(\ref{ortsphar}) can be solved as 
\bea
I
&=&\frac{\pi}{2}\sum_{l=0}^{\infty}\sum_{m=-l}^lA_{l,m}e^{-\lambda_{l,m}^2 t}
\delta_{k_{lm};k_{l'm'}}\frac{1}{k_{lm}^2} \delta_{ll'}\delta_{mm'}\nn.
\eea
Using Kronecker $\delta$'s to kill those orthogonal summations and continuing from $l\rightarrow 
l'$ and from $m\rightarrow m'$ we have
\bea
I&=&A_{l',m'}e^{-\lambda_{l',m'}^2 t}
\frac{\pi}{2}\frac{1}{k_{l'm'}^2} \nn.
\eea
These solutions are based upon the independence of time and now we introduce this with an exponential. So with the boundary condition of $t=0$ we find:
\bea
I_0&=&\int_Vj_{l'm'}({k_{l'm'}{r^\prime})}Y_{l'}^{m'}(\theta,\phi)\rho({r^\prime},\theta,\phi;0) d^3{r^\prime} \nnc
&=&A_{l',m'} 
\frac{\pi}{2}\frac{1}{k_{l'm'}^2}.
\eea
This leaves us with the final result of
\bea
A_{l,m}=\frac{2 k_{lm}^2}{\pi}\int_Vj_l({k_{lm}r)}Y_{l}^{m}(\theta,\phi)\rho({r^\prime},\theta,\phi;0) d^3{r^\prime}.
\eea

\end{document}